\documentclass[10pt]{iopart}
\usepackage{epsf}
\begin{document}
\newcommand{\vsp }{\vspace*{3mm}      }
\newcommand{\bay }{\begin{array}      }
\newcommand{\eay }{\end{array}        }
\newcommand{\slL }{\sum_{\ell=0}^L             }
\renewcommand{\Tr}{\mathop{\mbox{\rm Tr}}      }
\newcommand{\bsi }{{\mbox{\boldmath $\sigma $}}}
\newcommand{\ol  }{\overline    }
\newcommand{\fa  }{\forall      }
\newcommand{\nl  }{ \newline    }
\newcommand{\noi }{\noindent    }
\newcommand{\nn  }{\nonumber    }
\newcommand{\ov  }{\over        }
\newcommand{\hsp }{\hspace*     }
\newcommand{\cF  }{{\cal   F }  }
\newcommand{\cH  }{{\cal   H }  }
\newcommand{\cK  }{{\cal   K }  }
\newcommand{\cN  }{{\cal   N }  }
\newcommand{\cZ  }{{\cal   Z }  }
\newcommand{\lgl }{\langle      }
\newcommand{\rgl }{\rangle      }
\newcommand{\lv  }{\left[       }
\newcommand{\rv  }{\right]      }
\newcommand{\lh  }{\left(       }
\newcommand{\rh  }{\right)      }
\newcommand{\lc  }{\left\{      }
\newcommand{\rc  }{\right\}     }
\newcommand{\rp  }{\right .     }
\newcommand{\si  }{\sigma       }
\newcommand{\be  }{\beta        }
\newcommand{\de  }{\delta       }
\newcommand{\ep  }{\epsilon     }
\newcommand{\ta  }{\tau         }
\newcommand{\bx  }{{\bf x      }}
\newcommand{\tim }{{\tilde{ m }}}
\newcommand{\tin }{{\tilde{ n }}}
\newcommand{\ev  }{\equiv       }
\renewcommand{\e }{\!+\!        }
\newcommand{\m   }{\!-\!        }
\jl{1}

\letter{Cluster derivation of Parisi's RSB solution \\ 
        for disordered systems}

\author{J.~van~Mourik and A.C.C.~Coolen}

\address{ Department of Mathematics, King's College London,
          The Strand, London WC2R~2LS, UK}
\date{\today}

\begin{abstract}
We propose a general scheme in which disordered systems are allowed to
sacrifice energy equi-partitioning and separate into a hierarchy of ergodic
sub-systems (clusters) with different characteristic time-scales and
temperatures.
The details of the break-up follow from the requirement of stationarity of the 
entropy of the slower cluster, at every level in the hierarchy.
We apply our ideas to the Sherrington-Kirkpatrick model, and show how the Parisi
solution can be {\it derived} quantitatively from plausible physical principles.
Our approach gives new insight into the physics behind Parisi's solution and its
relations with other theories, numerical experiments, and short range models.
\end{abstract}


\hsp{5mm}

\noi
The Parisi scheme \cite{P} for replica symmetry breaking (RSB) has been one of
the most celebrated tools in the description of the `glassy' phase of disordered
systems.
It was initially proposed as the solution for the Sherrington-Kirkpatrick (SK)
model \cite{SK} for spin glasses, but it has since then been successfully 
applied to a wide range of models.
The physical interpretation of Parisi's solution has been the subject of many
discussions, and has generated notions such as hierarchies of disparate
time-scales \cite{S}, effective temperatures \cite{BP}, low entropy production
\cite{CK} and  non-equilibrium thermodynamics \cite{N,Franz}.
Central is the idea of multiple temperatures, which are usually defined via the
violation of fluctuation-dissipation relations; this often limits studies to
very specific models where correlation- and response functions can be calculated
explicitly.
In this letter, in contrast, we present and derive a general scheme in which
disordered systems are allowed to sacrifice full energy equi-partitioning by
separating autonomously into a hierarchy of ergodic sub-systems with different
characteristic time-scales; the statistics at every level (including effective
temperatures) follow from the ${\cal H}$-theorem with constrained (i.e. {\em
stationary}) entropy.
When applied to the SK model, our scheme is found to yield the Parisi solution
and to generate and connect the above concepts in a transparent way.
Our assumptions are simple and natural, and all ingredients of our theory have
a clear physical meaning.
Our study proceeds in three distinct stages.
First we show generally how and why multiple temperatures can arise in
disordered systems.
We then show how this generates replica theories with nested levels of
replication, with dimensions reflecting ratios of temperatures.
We apply our ideas to the `benchmark' disordered system, the SK model, and
{\em derive} Parisi's solution.
We close this letter with numerical evidence for the existence of multiple
disparate time-scales, a summary of the simple physical picture that naturally
emerges from our scheme, and a discussion of the points which need further
investigation.
\vsp

To understand the origin of multiple temperatures in a system of stochastic
variables $\bsi=(\sigma_1,\ldots,\sigma_N)$ with Hamiltonian $H(\bsi)$ and state
probabilities (or densities) $p(\bsi)$, we turn to Boltzmann's
${\cal H}$-function ${\cal H}=\Tr_{\bsi} p(\bsi)\lc H(\bsi)+T\log p(\bsi)\rc $,
which decreases monotonically under standard Glauber or Langevin dynamics and is
bounded from below by the free energy of the Boltzmann state.
For the case where we have two groups of variables (fast vs. slow), i.e.
$\bsi=(\bsi_{\rm f},\bsi_{\rm s})$, we substitute $p(\bsi_{\rm f},\bsi_{\rm s})=
p(\bsi_{\rm f}|\bsi_{\rm s}) p(\bsi_{\rm s})$ and find
\begin{eqnarray}
{\cal H}=\Tr_{\bsi_{\rm s}}~p(\bsi_{\rm s})\lc H_{\rm eff}(\bsi_{\rm s})+
T\log p(\bsi_{\rm s})\rc 
\label{eq:slow-Hfunction} \\
H_{\rm eff}(\bsi_{\rm s})=\Tr_{\bsi_{\rm f}}~p(\bsi_{\rm f}|\bsi_{\rm s})
\lc H(\bsi_{\rm f},\bsi_{\rm s})+T\log p(\bsi_{\rm f}|\bsi_{\rm s})\rc 
\label{eq:fast-Hfunction}
\end{eqnarray}
In the case where $\bsi_{\rm s}$ and $\bsi_{\rm f}$ evolve on disparate
time-scales, the minimisation of (\ref{eq:slow-Hfunction}) will occur in stages.
First, for every (fixed) $\bsi_{\rm s}$ the distribution
$p(\bsi_{\rm f}|\bsi_{\rm s})$ of the fast variables will evolve such as
to minimize (\ref{eq:fast-Hfunction}), i.e. towards the Boltzmann state
\begin{equation}
p(\bsi_{\rm f}|\bsi_{\rm s})=\cZ^{-1}_{\rm f}(\bsi_{\rm s})~
e^{-\be H(\bsi_{\rm f},\bsi_{\rm s})}
~~~~~~~
\cZ_{\rm f}(\bsi_{\rm s})=\Tr_{\bsi_{\rm f}}~
e^{-\be H(\bsi_{\rm f},\bsi_{\rm s})}
\label{eq:boltzmann_fast}
\end{equation}
Finding multiple temperatures requires, in addition to disparate time-scales,
stationarity of the entropy of the slow system (on the relevant `glassy'
time-scales).
Now (\ref{eq:slow-Hfunction}) is minimised subject to the constraint that the
entropy $S_{\rm s}=-\Tr_{\bsi_{\rm s}}~p(\bsi_{\rm s})\log
p(\bsi_{\rm s})$ be kept constant, giving
\begin{equation}
p(\bsi_{\rm s})=\cZ_{\rm s}^{-1}e^{-\tilde{\be }H_{\rm eff}(\bsi_{\rm s})}
~~~~~~~~
\cZ_{\rm s}=\Tr_{\bsi_{\rm s}}~e^{-\tilde{\be } H_{\rm eff}(\bsi_{\rm s})}
\label{eq:boltzmann_slow}
\end{equation}
i.e. a Boltzmann state for the slow variables, with the free energy of the fast
ones acting as effective Hamiltonian, and at inverse temperature
$\tilde{\be }=\tim \be $.
This leads to an $\tim$-dimensional replica theory, since combining
(\ref{eq:fast-Hfunction},\ref{eq:boltzmann_fast},\ref{eq:boltzmann_slow}) gives
$\cZ_{\rm s}=\Tr_{\bsi_{\rm s}}[\cZ_{\rm f}(\bsi_{\rm s})]^\tim$.
The dimension $\tim$ follows from demanding the prescribed value of the slow
entropy: $\be  \tim^2 (\partial \cF_{\rm s}/\partial \tim)=S_{\rm s}$, with
$\cF_{\rm s}=-\tilde{\be }^{-1}\log \cZ_{\rm s}$.
For $T>\tilde{T}$ the fast variables would start acting as a heat bath for the
slow ones, so thermodynamic stability requires $\tim\leq 1$.
Note that $\tim <1$ implies that the contraining entropy must be larger than
that of the Boltzmann state (indeed, a large characteristic time scale does not
imply low entropy).

The above argument can be generalised to an arbitrary hierarchy.
The variables $\bsi_\ell$ at each level $\ell$ are characterised by distinct
time-scales and temperatures $\{\tau_\ell,\be _\ell\}$ ($\ell=0,1,\ldots,L$);
each level being adiabatically slower than the next,
$\tau_\ell\ll \tau_{\ell\m 1}$.
This leads to replicating recursion relations for the partition sums
at subsequent levels:
\begin{equation}\hsp{-1.5mm}\bay{l}
\cZ_\ell=\Tr_{\bsi_\ell}\left[\cZ _{\ell+1}\right]^{\tim_{\ell+1}}
~~~~~~~~~~~~~~~(\ell<L) \\[2mm]
\cZ_L   =\Tr_{\bsi_L}~e^{-\be_L H(\{\bsi \})}\eay
\label{dt:Zl}
\end{equation}
with $\tim_ \ell= \be _{\ell-1}/\be _\ell\leq 1$, and $\be_L=\be $.
The replica dimensions $\tim_\ell$ follow from the prescribed (stationary, but
as yet unkwown) values $S_\ell$ of the level-$\ell$ entropies, via $\be
_{\ell+1}\tim_{\ell+1}^2 (\partial\cF_\ell/\partial \tim_{\ell+1})=S_{\ell}$,
with $\cF_{\ell}=-\be _{\ell}^{-1}\log \cZ_{\ell}$.
Equivalently, using the specific nesting of the partition functions in
(\ref{dt:Zl}) one shows that the $\{\tim_\ell\}$ are uniquely determined by the
identities
\begin{equation}
\be _{\ell+1}\tim_{\ell+1}^2 \frac{\partial}{\partial
\tim_{\ell+1}}\cF_{0}=\Sigma_\ell
~~~~~~~~~~
\Sigma_\ell=\lgl~\lgl\ldots\lgl~S_{\ell}~\rgl_{\ell-1}\ldots\rgl_1\rgl_0
\label{eq:finding_m}
\end{equation}
in which $\lgl \cdots\rgl _r$ denotes the average over the equilibrated
level-$r$ process.
Due to the constrained minimisations underlying (\ref{dt:Zl}), the free
energies $\cF_\ell$ are generally not minimised; however, one can verify that
$\cF_0$ still serves as a generator of observables:
\begin{equation}
H(\{\bsi\})\to H(\{\bsi\})+\lambda \psi(\{\bsi\}):~~~~~~~
\lim_{\lambda\to 0}\frac{\partial}{\partial\lambda}\cF_0=\lgl \psi(\{\bsi\})\rgl
\label{eq:manylevel_generator}
\end{equation}
This generalises a formalism originally developed and applied for spin systems
with slowly evolving bonds \cite{PC}.
The construction reverts back to the conventional statistical mechanical
picture if the constraining entropies $S_\ell$ are identical to those of the
full Boltzmann state: then the constraining forces vanish and $\tim_\ell=1$ for
all $\ell$.
\vsp

We now apply this scheme to the SK model \cite{SK}, for which the Parisi
solution was originally constructed, which describes $N$ Ising spins
with the conventional Hamiltonian
$H(\bsi )=-\sum_{i<j} J_{ij}\si _i\si _j$,
but with suitably scaled independent random couplings $J_{ij}$
(with average $J_0/N$ and variance $J/\sqrt{N}$).
We assume, following our previous arguments, that this system can be viewed as
a hierarchy of $L\e 1$ levels of spins, each level $\ell$ with distinct
disparate time-scales and temperatures $\{\tau_\ell,T_\ell\}$:
\begin{equation}
\{1,\ldots,N\}=\bigcup_{\ell=0}^L I_\ell,~~~~~
\bsi=(\bsi_0,\ldots,\bsi_L),~~~~~
\bsi_\ell=\{\si_j|j\in I_\ell\}
\end{equation}
with $|I_\ell|=\ep_\ell N$, and such that $\ta _\ell\ll\ta _{\ell\m 1}$ for all
$\ell$ (thus larger values of $\ell$ correspond to faster spins).
The selection of time-scales for the spins is expected to depend on the
realisation of the couplings, but here we will make the simplest approximation:
the system can only choose the relative level sizes $\{\ep _\ell\}$.
A study of the autonomous selection of levels will be presented elsewhere
\cite{new}.
We calculate the disorder-averaged free energy $\cF_0$ (the general multi-level
generator of observables) with the replica trick
\begin{equation}
\ol {\cF_0 }=-\be_0^{-1}\ol {\log\cZ _0}=-\lim_{\tin \to 0}~(\tin \be
_0)^{-1}
\log\ol {\cZ _0^\tin }
\label{sk:F}
\end{equation}
\noi Together with the relations (\ref{dt:Zl}), this leads us to a nested
set of $\tin \prod_{\ell=1}^L\tim _\ell$ replicas. A spin at level
$\ell$ thus carries a set $\{a\}_\ell= \{a_0,..,a_\ell\}$ of replica indices,
where $a_0\in\{1,..,\tin\}$ reflects the disorder average, and
with $a_\ell\in\{1,..,\tim _\ell\}$.
As before $\tim_\ell=\be _{\ell-1}/\be _\ell\leq 1$.
Following standard manipulations, the asymptotic free energy per spin
$f=\lim_{N\to\infty}\ol {\cF_0}/N$ is then found to be
\begin{eqnarray}
\hsp{-1cm}
f&=&\lim_{\tin \to 0}~{1\ov \tin \be _0}~{\rm extr}
\lv \frac{J^2\be^2}{4}\!\!\sum_{\{a\}_L,\{b\}_L}\! q^{\{a\}_L~2}_{\{b\}_L}-
\slL ~\ep _\ell~\log ~\cK _\ell\rv
\label{sk:fab}\\
\hsp{-1cm}
\cK _\ell&= &\!\Tr_{\{\si ^{\{c\}_\ell }\}}\exp\lv \frac{J^2\be^2}{2}\!\!\!
\sum_{\{a\}_L, \{b\}_L}\!q^{\{a\}_L}_{\{b\}_L}~\si ^{\{a\}_\ell}
\si ^{\{b\}_\ell}\rv
\label{sk:qab}
\end{eqnarray}
Extremisation is to be carried out with respect to the order parameters
$q^{\{a\}_L}_{\{b\}_L}$, whose physical meaning is given by (with averages
denoting the multi-temperature statistics):
\begin{equation}
q^{\{a\}_L}_{\{b\}_L}=\lim_{N\to\infty} {1\ov N}\sum_{\ell=0}^L
\sum_{j\in I_\ell}\lgl \si _j^{\{a\}_\ell}\si _j^{\{b\}_\ell}\rgl
\end{equation}
With the new definitions $m_\ell= \prod_{k=\ell}^L \tim_k=\be _{\ell-1}/\be$
we obtain $\be_0\tin =\be n$, and the connection with the original Parisi
solution becomes clear.
What remains is to assume full ergodicity within each level in the hierarchy of
time-scales:
\begin{equation}
q^{\{a\}_L}_{\{b\}_L}=q_{\ell[\{a\}_L,\{b\}_L]}
\label{sk:qrs}
\end{equation}
where $\ell[\{a\}_L,\{b\}_L]$ denotes the slowest level for which the the two
strings of replica coordinates $\{a\}_L$ and $\{b\}_L$ differ.
Insertion of (\ref{sk:qrs}) into (\ref{sk:fab}) gives
\begin{eqnarray}
f=\frac{\be  J^2}{2}\slL \lv \frac{1}{2}m_{\ell\e 1}(q^2_{\ell\e 1}-
q_\ell ^2)-\ep _\ell \sum_{r=\ell}^{L}m_{r\e 1}(q_{r\e 1}-q_r) \rv
\nn \\\hsp{40mm}
-~{1\ov m_1\be }\slL \ep_\ell\int\! Dz_0~\log[\cN ^1_\ell]
\label{sk:frs}
\end{eqnarray}
\vspace*{-2mm}
where
\begin{equation}
   \cN ^r_\ell  = \lc \bay {lll}
\int Dz_r [\cN ^{r+1}_\ell]^{m_r\ov m_{r\e 1}}
&{\rm for}& r\leq \ell \\
        2\cosh(J\be m_{\ell\e 1}\sum_{s=0}^\ell z_s\sqrt{q_s\m q_{s\m 1}})
&{\rm for}& r=\ell+1         \eay \rp
\label{sk:N}
\end{equation}
The physical meaning of $q_\ell$ is
\begin{equation}
q_\ell=\lim_{N\to\infty}\frac{1}{N}\sum_j\ol {\lgl ~..~\lgl ~\lgl ~\lgl ~..~\lgl
\si _j\rgl _L~..~\rgl _{\ell\e1} ~\rgl _\ell^2 ~\rgl _{\ell\m 1}~..~\rgl _0}
\label{ski:qm}
\end{equation}
in which $\ol {\cdots}$ denotes the disorder average.
The physical saddle-point is the analytic continuation of the one which
minimises $f$ for positive integer values of $\{\tin , \tim _\ell\}$. For such
values, the minimum with respect to the $\ep _\ell$ (with
$\sum_{\ell=0}^L\ep _\ell =1$), in turn, is found to occur for
$\{\ep^*_L=1,~\ep ^*_\ell=0~~\fa \ell<L\}$, i.e. in the thermodynamic limit the
slow spins form a vanishing fraction of the system as a whole.
We have now exactly recovered the $L$-th order Parisi solution.
The values of the $m_\ell$ follow from (\ref{eq:finding_m}), which translates
into
\begin{equation}
\be m_{\ell+1}^2 \frac{\partial}{\partial m_{\ell+1}}f=\ol {\Sigma}_\ell/N
\label{eq:finding_m_2}
\end{equation}
The bounds $0\leq \lim_{N\to\infty}\Sigma_\ell/N\leq \epsilon_\ell\log 2$
subsequently dictate that, as $\epsilon_\ell\to 0$ for all $\ell<L$,
determining $m_\ell$ via (\ref{eq:finding_m_2}) simply reduces to extremising
$f$ with respect to $m _\ell$, thus removing the need to know the values of the
constraining entropies $S_\ell$.

\begin{figure}[t]
\vspace*{0.5cm}
\setlength{\unitlength}{0.6mm}
\begin{picture}(-1,90)
\put( 65, 10){\epsfysize=80\unitlength\epsfbox{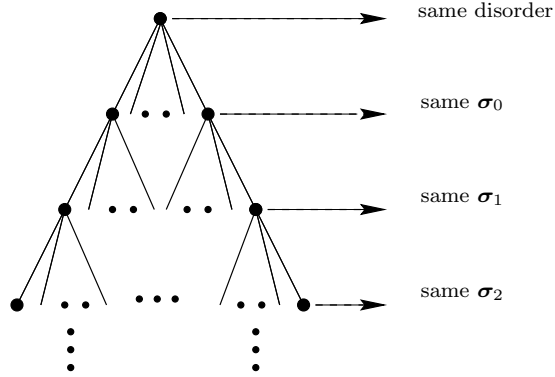}}
\put(170, 90){\makebox(3,1)
{\footnotesize same disorder                                }}
\put(164, 69){\makebox(3,1)
{\footnotesize same {\boldmath $\sigma$}$_0$}}
\put(164, 48){\makebox(3,1)
{\footnotesize same {\boldmath $\sigma$}$_1$}}
\put(164, 27){\makebox(3,1)
{\footnotesize same {\boldmath $\sigma$}$_2$}}
\end{picture}
\vspace*{-7mm}
\caption{The ultra-metric tree, which here is a direct consequence of the
hierarchy of spin clusters, evolving at disparate time-scales.}
\label{figUl}
\end{figure}

We have thus shown that the Parisi solution can be derived from simple physical
principles, and can be interpreted as describing a system with an infinite
hierarchy of time-scales where a vanishingly small fraction of slow spins act
as effective symmetry-breaking disorder for the faster ones.
The vanishing of the fraction of slow spins indicates that the cumulative
entropy of the slow spins is sub-extensive, and that the so-called {\it
complexity} is zero.
A block-size  $m _\ell$ at level $\ell$ of the Parisi matrix is found to be the
ratio of the effective temperature $T_\ell$ of that level and the ambient
temperature $T$.
Extremization of the free energy per spin with respect to $m_\ell$ is equivalent
to saying that the average entropy of the spins at level $\ell-1$ is stationary
and sub-extensive.
It follows from physical considerations (no heat flow in equilibrium) that
$m_\ell\leq 1$ for all $\ell$.
Ultra-metricity (see fig.~\ref{figUl}) is a direct consequence of the existence
of a hierarchy of time-scales.
At each level $\ell$, the different descendants of a node represent different
configurations of the $\bsi _{\ell\e 1}$, which share the same realisation of
the disorder and of the slower spins.

\begin{figure}[t]
\vspace*{11mm}
\setlength{\unitlength}{0.7mm}
\begin{picture}(0,190)
\put( 45,100){\epsfxsize=150\unitlength\epsfbox{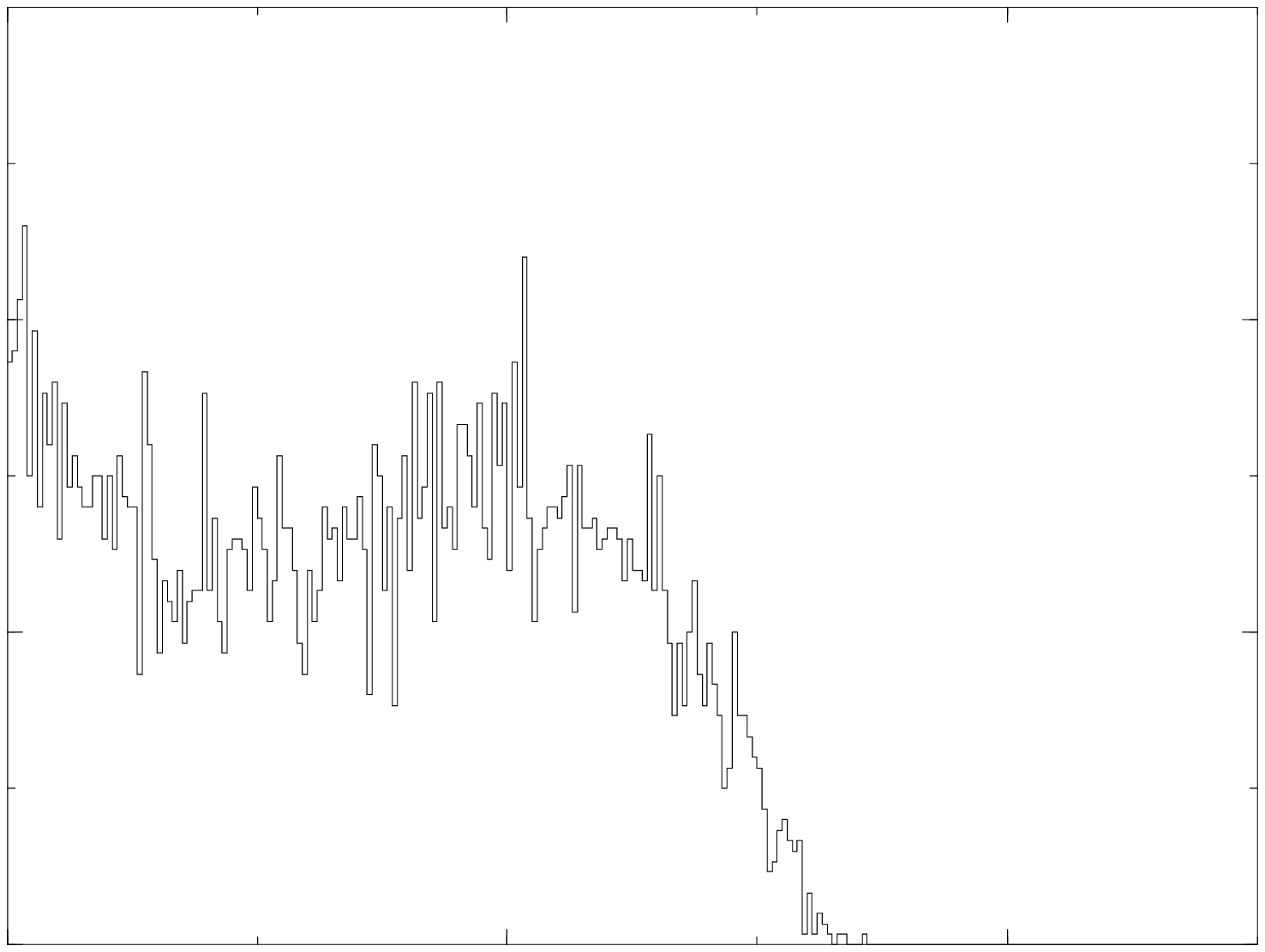}}
\put(120,104){\makebox(3,1){$f$}}
\put(69,111){\scriptsize $0$}
\put(109,111){\scriptsize $10^5$}
\put(154,111){\scriptsize $2.10^5$}
\put( 36,160){\makebox(3,1){$\rho_{\rm sim}(f)$}}
\put( 56,119){\makebox(3,1){\scriptsize $0$}}
\put( 56,146){\makebox(3,1){\scriptsize $0.5~10^{-5}$}}
\put( 56,173){\makebox(3,1){\scriptsize $1.0~10^{-5}$}}
\put( 56,200){\makebox(3,1){\scriptsize $1.5~10^{-5}$}}
\put( 69, 20){\epsfxsize=112\unitlength\epsfbox{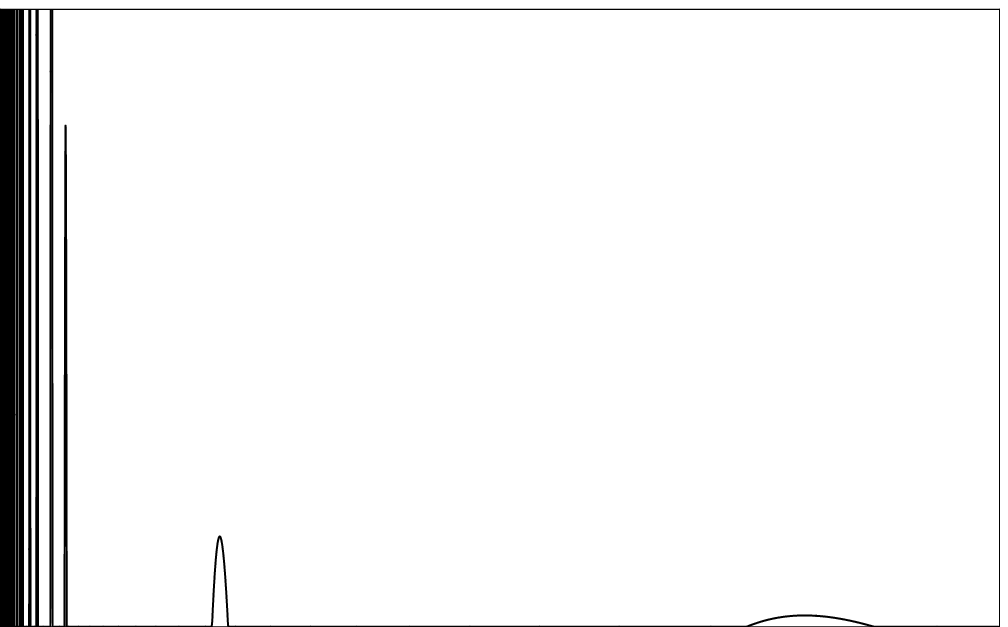}}
\put( 60,88){\makebox(3,1){\scriptsize $10^{-6}$}}
\put( 45, 58){\makebox(3,1){$W^*(\tau)$}}
\put( 60, 20){\makebox(3,1){\scriptsize $0$}}
\put( 68, 14){\makebox(3,1){\scriptsize $0$}}
\put(120, 10){\makebox(3,1){\large $\tau$}}
\put(180, 14){\makebox(3,1){\scriptsize $1.5~10^{5}$}}
\put( 115, 35){\epsfxsize= 60\unitlength\epsfbox{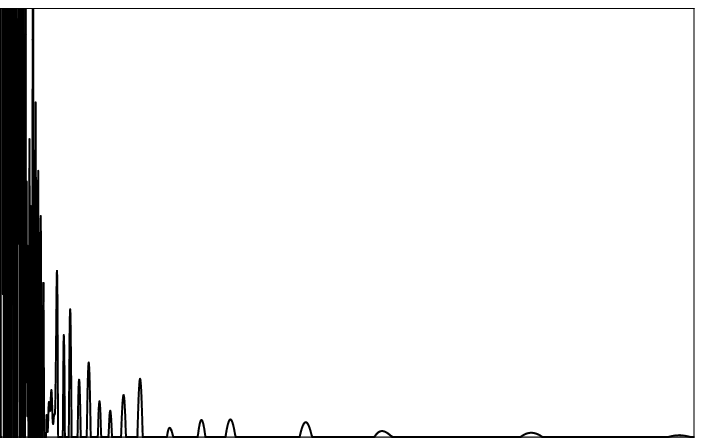}}
\put( 106,83){\makebox(3,1){\scriptsize $2~10^{-5}$}}
\put( 106, 46){\makebox(3,1){\scriptsize $0$}}
\put( 114, 41){\makebox(3,1){\scriptsize $0$}}
\put(174, 41){\makebox(3,1){\scriptsize $10^{4}$}}
\end{picture}
\vspace*{-6mm}
\caption{Upper graph: distribution of the number of flips $f$ per spin for a
simulation of the SK-model with $N=6000$, during $t=10^6$ Monte-Carlo updates
per spin, after a waiting time of $t_w=6.10^5$, at $T=0.25$. Lower graph:
corresponding estimate of the most probable distribution $W^\star(\tau)$ of
time-scales $\tau$. Inset: the small $\tau$ area enlarged.}
\label{figft}
\end{figure}

Since our proposal relies fundamentally on the existence of clusters with
widely separated characteristic time-scales, we sought to provide independent
evidence for this assumption by measuring the distribution
$\rho_{\rm sim}(f,t)$ of the number of flips $f$ per spin at time $t$ in
numerical simulations of the SK-model, see fig.~\ref{figft}.
Upon assuming an independent characteristic time-scale $\tau_j$ for each spin
$\si _j$, and a distribution $W(\tau)$ for these time-scales, one obtains a
simple theoretical prediction for this distribution:
\begin{equation}
\rho_{\rm th}(f,t|W)\simeq\int_0^\infty\! d\tau~W(\tau)
\lh \!\!\bay {c}t\\f\eay \!\!\rh {1\ov \tau^f}~(1\m {1\ov \tau})^{t-f}
\end{equation}
Minimising the deviation $\sum_{f=0}^\infty\lv \rho_{\rm sim}(f,t)-\rho_{\rm th}
(f,t|W)\rv ^2$ with  respect to the $W(\tau)$ yields an estimate of the most
probable distribution of time-scales $W^*(\tau)$, see fig.~\ref{figft}, which
clearly supports our assumptions.
Both the number of peaks (in agreement with full RSB), and the separation
between the peaks (in agreement with infinitely disparate time-scales) are
found to grow with increasing system size and/or time, whereas the fraction of
`slow' spins appears to decrease with increasing system size.
\vsp

\begin{figure}[t]
\setlength{\unitlength}{0.6mm}
\begin{picture}(0,120)
\put( 80,10){\epsfxsize=150\unitlength\epsfbox{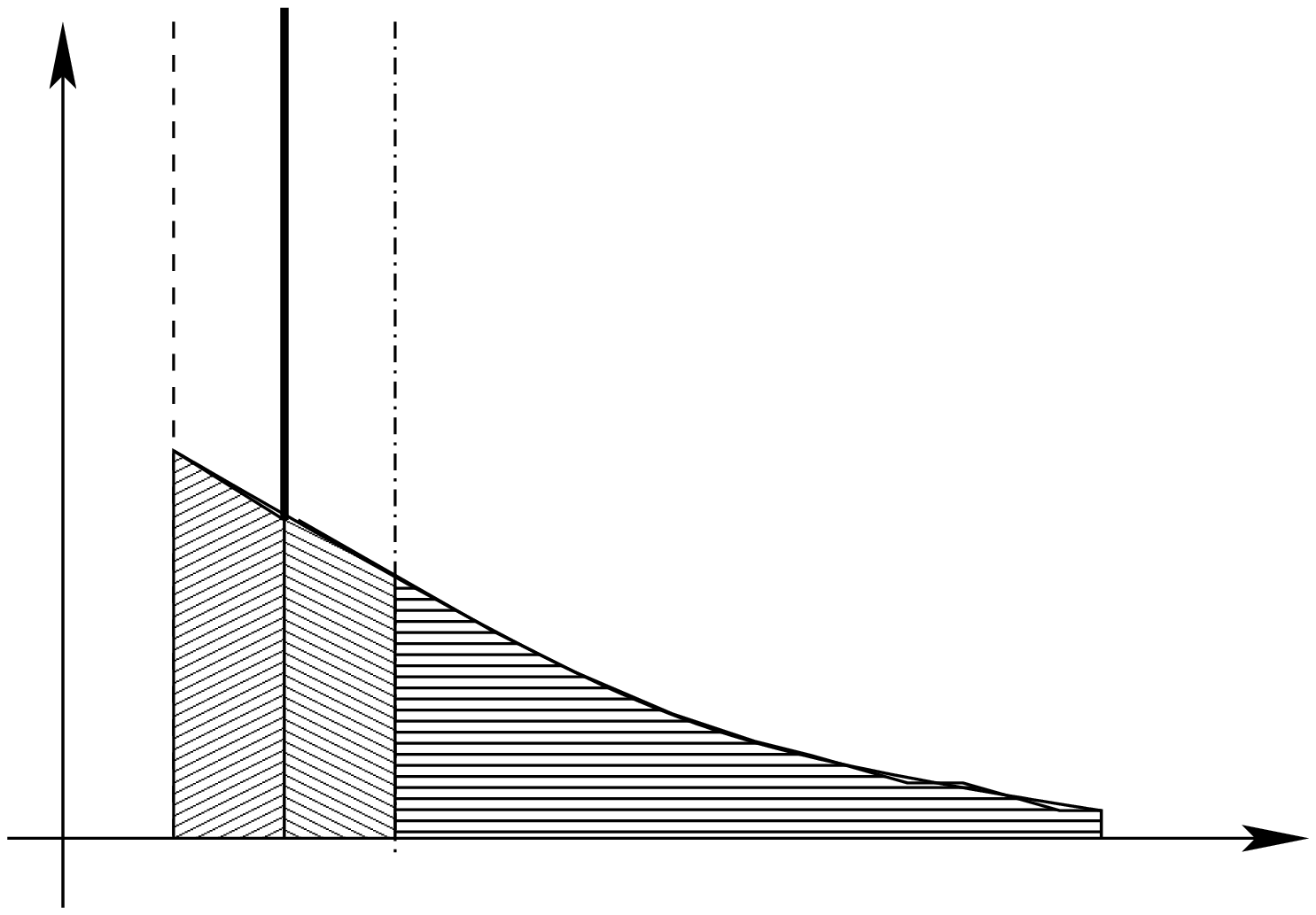}}
\put(160,8){$T_{\rm eff}$}
\put(97,8){$T_1$}
\put(110,8){$T$}
\put(123,8){$T_2$}
\put(55,75){$W(T_{\rm eff})$}
\end{picture}
\vspace*{-3mm}
\caption{Qualitative sketch of the distribution $W(T_{\rm eff})$ of (effective)
temperatures (note: time-scales increase with $T_{\rm eff}$) at ambient
temperature $T$ (resp. $T_1$, $T_2$) in the spin glass phase.}
\label{figTe}
\end{figure}

In fig.~\ref{figTe} we sketch the qualitative picture emerging from our
interpretation of the Parisi scheme. Most spins  evolve at
the fastest (microscopic) time-scale, at ambient temperature $T$;  a
small fraction evolves at (infinitely) slower time-scales, at higher
effective temperatures. Cooling to a temperature $T_1\!<\!T$, followed by
heating back to $T$, will leave spins with $T_{\rm eff}\!>\!T$ unchanged,
explaining memory effects. Conversely, after heating to $T_2\!>\!T$ and
cooling back to $T$, the original states of spins with $T\!\leq\!
T_{\rm eff}\!\leq\!T_2$ will be erased, which may explain thermo-cycling
experiments (for a recent review see e.g. \cite{TC}).
We expect the qualitative features of our picture to survive in short range
systems, where the time-scales need not be infinitely disparate due to
activated processes.
The origin of the slow time-scales  of these clusters must lie in the latter
being coupled much stronger internally, than (effectively) to the rest of the
system.
They could therefore be seen as a `soft' version of the fully disconnected
clusters which give rise to so-called Griffiths singularities in diluted
systems \cite{G}.
In short range systems, the clusters would have to be spatially localized, in
line with the {\it droplet} picture proposed by Fisher and Huse \cite{FH}.
In such systems, each of the different levels would correspond to multiple
localised spin clusters.
The fact that the characteristic time-scale of a cluster increases with
$T_{\rm eff}\m T$ explains why the effective age of a system at temperature $T$
is found to decrease upon spending time at $T_1<T$, but to increase upon doing
so at $T_2>T$.

At a theoretical level,
a more careful treatment of the selection of clusters is clearly needed (and
is currently being carried out \cite{new}), both for full- and 1-RSB models.
This may allow us to calculate the complexity in such systems. Furthermore, it
needs to be investigated whether slow clusters survive above the thermodynamic
spin-glass temperature $T_{\rm sg}$. Our results also suggest further numerical
experiments for both mean field and short range models, concentrating on
quantities such as spin flip frequencies, avalanches, spatial correlations, and
cluster persistency \cite{KH,BZ,RZ}.
\vsp

\noi
It is our pleasure to thank F.~Ritort and D.~Sherrington for critical comments
and stimulating discussions.
\vsp


\end{document}